\documentclass[aps,prd,reprint,twocolumn]{revtex4-1}
\usepackage{amsmath}
\usepackage{amssymb}
\usepackage{graphicx}
\usepackage{times}
\usepackage{euscript}
\usepackage{amsthm}
\usepackage{amsfonts}
\usepackage[english]{babel}
\usepackage{epstopdf}
\usepackage{multirow,makecell,array}
\usepackage{mathtools}
\usepackage[dvipsnames]{xcolor}
\newcommand{\fdiag}[3][0pt]{%
  \vcenter{\hbox{\raisebox{#1}{\includegraphics[height=#2]{#3}}}}%
}

\newcommand{\smallleftrightarrow}{\mathrel{\leftrightarrow}}
\newcommand{\darrow}[1]{\overset{\scriptscriptstyle\smallleftrightarrow}{#1}}

\makeatletter
\newcommand{\varoverset}[2]{\binrel@{#2}%
  \binrel@@{\mathop{\vphantom{M}{#2}}\limits^{#1}}}
\makeatother

\begin{document}

\title{Vacuum photon emission and mean electromagnetic field in pair-creating external backgrounds}

\author{I.~A.~Aleksandrov}
\author{E.~V.~Perelygin}
\author{D.~V.~Chubukov}
\affiliation{Department of Physics, Saint Petersburg State University, Universitetskaya Naberezhnaya 7/9, Saint Petersburg 199034, Russia}

\begin{abstract}
We develop a perturbative description of vacuum radiative processes in quantum electrodynamics with a prescribed external electromagnetic background capable of producing electron-positron pairs. Since the initial vacuum is then unstable and the in- and out-vacua are inequivalent, radiative observables require a real-time formulation beyond the ordinary in-out approach of vacuum-stable QED. Using the Keldysh-Schwinger-Fradkin nonequilibrium technique, we derive the mean number density of emitted photons through the second nonvanishing order in the fine-structure constant. The leading term, of order $\alpha$, reproduces the known vertex and tadpole mechanisms, while the complete order-$\alpha^2$ correction contains interference, loop, and induced-current contributions. We also give an independent derivation based on the spectral decomposition of the identity operator in the in-Fock space, where the photon number density is represented as a sum of squared transition amplitudes and vacuum-disconnected terms are canceled by the optical theorem generalized to an unstable vacuum. In addition, we compute the mean electromagnetic field through order $e^3$, including the electromagnetic dressing of the induced vacuum current, and verify it using the corresponding Schwinger-Dyson equations. The final formulas are expressed in terms of exact solutions and propagators of the Dirac equation in the external background and apply to general spacetime-dependent field configurations.
\end{abstract}

\maketitle


\section{Introduction}\label{sec:intro}

One of the central lessons of quantum electrodynamics (QED) is that the vacuum is a polarizable quantum medium. In external electromagnetic fields, virtual electron-positron pairs generate nonlinear corrections to Maxwell electrodynamics, encoded at low energies in the one-loop Heisenberg-Euler effective action~\cite{euler_kockel,heisenberg_euler}. Purely magnetic backgrounds preserve vacuum stability, whereas Schwinger's proper-time analysis showed that a constant electric field gives the effective action a nonzero imaginary part~\cite{schwinger_1951}, signaling the decay of the initial vacuum into states with real electron-positron pairs. QED in a pair-creating background therefore differs qualitatively from vacuum-stable perturbation theory: the in- and out-vacua are inequivalent, vacuum persistence is nontrivial, and expectation values require a real-time formulation adapted to an unstable vacuum.

External classical fields provide a natural framework for describing strong-field QED phenomena. In many important cases the background does not lead to vacuum decay; for example, a plane electromagnetic wave cannot produce pairs from the vacuum, independently of its intensity~\cite{nikishov_ritus_1964,brown_kibble_1964}. More general configurations may violate vacuum stability, typically near the critical scale $E_\mathrm{c} = m^2 c^3/|e\hbar| \sim 10^{16}~\mathrm{V/cm}$, where $m$ and $e$ are the electron mass and charge. Although this scale is far above ordinary laboratory electric fields, high-power laser facilities are steadily bringing the nonlinear strong-field regime within experimental reach~\cite{dipiazza_rmp_2012,gonoskov_2022,fedotov_review_2023,popruzhenko_ufn_2023,kostyukov_fian_2024}. A complementary route is offered by collisions of highly charged heavy nuclei, where supercritical Coulomb fields may trigger spontaneous positron production~\cite{greiner_sfqed,maltsev_prl_2019,popov_prd_2020}. These developments motivate a systematic formulation of QED observables in pair-creating external backgrounds.

Radiative processes provide direct probes of charged-particle dynamics and vacuum fluctuations in strong external fields. In a pair-creating background, the mean photon number density, resolved with respect to momentum and polarization, receives two qualitatively different leading-order contributions~\cite{fradkin_gitman_shvartsman}. The first is the vertex channel, in which photon emission accompanies the creation of real electron-positron pairs~\cite{fradkin_gitman_shvartsman,otto_prd_2017,di_piazza_prd_2005,aleksandrov_prd_2019_2,aleksandrov_antonino_2022}. The second is the tadpole channel, in which radiation is generated by the induced vacuum current~\cite{fradkin_gitman_shvartsman,di_piazza_prd_2005,aleksandrov_prd_2019_2,fedotov_pla_2006,karbstein_prd_2015,gies_prd_2018_1,king_pra_2018,karbstein_prl_2019_2,aleksandrov_pra_2021}. This current contains contributions associated with real pair creation and vacuum polarization, and the corresponding tadpole term belongs to the broader class of external-field tadpole or one-particle-reducible contributions, which are known to require special care beyond the usual vacuum-QED intuition~\cite{gies_jhep_2017,karbstein_jhep_2017,ahmadiniaz_jhep_2019}. These order-$\alpha$ results are well understood and form the basis for the perturbative description of vacuum radiation in external backgrounds.

A systematic theory of radiative processes requires going beyond the leading contribution. Already at next-to-leading order, the photon number density involves interference terms, loop corrections, and contributions whose cancellation is tied to vacuum persistence. In contrast to stable-vacuum QED, disconnected vacuum loops in a pair-creating background are not hidden by the usual normalization of the scattering matrix and must be treated explicitly. Thus, even for observables with a transparent particle interpretation, a direct extension of textbook perturbation theory is not sufficient. The natural framework for real-time expectation values is the nonequilibrium, or closed-time-path, formalism of Schwinger and Keldysh~\cite{schwinger_ctp,keldysh_1965}. For QED with external classical fields, this approach was developed in a form particularly suited to unstable-vacuum problems by Fradkin and collaborators~\cite{fradkin_1981,fradkin_gitman_shvartsman}. This framework uses matrix propagators, and the normalization by the vacuum persistence amplitude is built in, so vacuum-disconnected contributions cancel automatically. Related work has shown that loop corrections in strong electric backgrounds can display secular dependence on the duration of a pair-creating field and may require resummation in sufficiently long pulses~\cite{akhmedov_jhep_2014,akhmedov_prd_2023}. Complementary coherent-state and worldline approaches have addressed backreaction, disconnected diagrams, and in-in observables in pair-creating backgrounds~\cite{copinger_prd_2025,copinger_arxiv_2025}.

In this work we use the Keldysh-Schwinger-Fradkin nonequilibrium formalism to derive the photon number density in a general external electromagnetic background through the second nonvanishing order in the interaction. The leading term, of order $e^2$ or $\alpha$, reproduces the known vertex and tadpole contributions. The main new result is the complete correction of order $e^4$, or $\alpha^2$, obtained in a closed form suitable for arbitrary spacetime-dependent backgrounds. The final expression is written in terms of exact fermionic modes and propagators in the external field and ordinary photon Green's functions.

We also present an independent derivation of the photon number density based on the spectral decomposition of the identity operator in the in-Fock space. This direct approach represents the observable as a sum of squared transition amplitudes and thus exposes the physical content of the result in terms of standard Feynman-type building blocks. Unlike the nonequilibrium calculation, it avoids the doubled matrix structure but generates vacuum-disconnected loops explicitly; these terms must be canceled by the optical theorem generalized to an unstable vacuum. The agreement of the two derivations provides a nontrivial check of the final result. At the same time, the direct method is specific to quantities with a particle-number interpretation and does not naturally extend to observables such as the mean electromagnetic field.

The mean electromagnetic field is the second observable considered in this paper. It contains both radiative and nonradiative components of the electromagnetic response, including vacuum-polarization effects and contributions associated with real particles produced from the vacuum. Its first nonvanishing term, proportional to $e$, is generated by the leading induced current and is known from previous studies~\cite{fradkin_gitman_shvartsman}. We derive the next correction, of order $e^3$, within the same Keldysh-Schwinger-Fradkin framework. This term describes the electromagnetic dressing of the induced current and cannot be obtained by the spectral-decomposition method used for the photon number density. It therefore illustrates the necessity of the nonequilibrium formalism for computing general expectation values in external-field QED with unstable vacuum.

As an additional consistency check, we compare the perturbative mean-field result with the corresponding Schwinger-Dyson equations~\cite{schwinger_pnas_1951,dyson_pr_1949}. These equations relate the mean electromagnetic field to the exact Heisenberg electron and photon propagators. Expanding them in the electromagnetic coupling, we verify that the first- and third-order terms obtained from the nonequilibrium generating functional satisfy the expected Schwinger-Dyson structure. This check complements the direct amplitude derivation of the photon number density and confirms the internal consistency of the perturbative construction.

The paper is organized as follows. Section~\ref{sec:setup} introduces the notation and the Keldysh-Schwinger-Fradkin generating functional. In Sec.~\ref{sec:photon_density_ksf}, we derive the photon number density within the nonequilibrium technique and present the complete result through order $\alpha^2$. Section~\ref{sec:photon_density_direct} presents an independent derivation based on the spectral decomposition of the identity operator and discusses the cancellation of vacuum-disconnected loops. In Sec.~\ref{sec:mean_field}, we compute the mean electromagnetic field through order $e^3$, and Sec.~\ref{sec:conclusions} contains our conclusions. The Schwinger-Dyson consistency check and useful identities are given in the appendices. Throughout the paper we use units $\hbar=c=1$ and $\alpha=e^2/(4\pi)$. The metric tensor is $g^{\mu \nu} = \mathrm{diag} (1, -1, -1, -1)$, and the position four-vector is denoted by $x^\mu = (x^0, \mathbf{x})$.


\section{Setup} \label{sec:setup}

This section introduces the notation and basic ingredients used throughout the paper. We specify the setup and the main definitions, including the generating functional for expectation values and its perturbative expansion.


\subsection{Basic definitions}

We work in the Furry picture~\cite{fradkin_gitman_shvartsman,furry_1951}, where the external classical background $\mathcal{A}_\mu$ is included exactly in the unperturbed problem, while the interaction $j^\mu A_\mu$ is treated perturbatively. Here $j^\mu = (e/2) [\overline{\psi}, \gamma^\mu \psi]$ is the current operator of the electron-positron field, and $A_\mu$ denotes the quantized part of the electromagnetic field. The latter has the standard mode expansion:
\begin{equation}
A^\mu (x) = \sum_{\lambda=0}^3 \int \!\! d\mathbf{k} \, \Big [ c_{\mathbf{k} \lambda} f^\mu_{\mathbf{k}\lambda} (x) + c^\dagger_{\mathbf{k} \lambda} f^{\mu*}_{\mathbf{k} \lambda} (x) \Big ],
\label{eq:photon_field}
\end{equation}
where
\begin{equation}
f^\mu_{\mathbf{k}\lambda} (x) = \frac{1}{(2\pi)^{3/2}} \frac{1}{\sqrt{2k^0}} \, \mathrm{e}^{-ikx} \varepsilon^\mu (\mathbf{k}, \lambda), \quad k^0 = |\mathbf{k}|.
\label{eq:photon_wf}
\end{equation}
Here $\varepsilon^\mu (\mathbf{k}, \lambda)$ denotes the polarization four-vector. We assume that the field strength of the external background vanishes as $|x^0| \to \infty$. The electron-positron field operator can then be expanded in the basis of \emph{in-solutions} of the Dirac equation in this background:
\begin{equation}
\psi (x) = \sum_{n} \big [a_n(\mathrm{in}) \, {}_+ \varphi_n (x) + b^\dagger_n(\mathrm{in}) \, {}_- \varphi_n (x)\big ]. \label{eq:psi_x_in}
\end{equation}
The functions $\{{}_\pm \varphi_n (x) \}_n$ form a complete set and are fixed by their asymptotic behavior at $x^0 \to -\infty$: at early times they are eigenfunctions of the one-particle Dirac Hamiltonian with positive and negative eigenvalues, respectively. The time-independent operators in Eq.~\eqref{eq:psi_x_in} define the \emph{in-vacuum} state,
\begin{equation}
a_n (\mathrm{in}) |0_e, \mathrm{in} \rangle = b_n (\mathrm{in}) |0_e, \mathrm{in} \rangle = 0
\label{eq:in_vac_def}
\end{equation}
for all $n$. The full in-vacuum vector is $|0, \mathrm{in} \rangle = |0_e, \mathrm{in} \rangle \otimes |0_\gamma \rangle$, where $|0_\gamma \rangle$ is the photon vacuum. Analogously, fixing the late-time asymptotic behavior defines the \emph{out-solutions} ${}^\pm \varphi_n (x)$ and the corresponding operators $a_n(\mathrm{out})$ and $b_n(\mathrm{out})$. In a pair-creating external background, the out-vacuum is inequivalent to the in-vacuum, so in general $|\langle 0, \mathrm{out}|0,\mathrm{in}\rangle| \neq 1$.

In what follows, we will need the causal photon and fermion propagators
\begin{align}
D_{\mu \nu} (x, y) &= -i \langle 0_\gamma| \mathcal{T} A_\mu (x) A_\nu (y) | 0_\gamma \rangle \nonumber \\
{}&= \theta (x^0 - y^0) D^{(-)}_{\mu \nu} (x, y) \nonumber \\
{}&- \theta (y^0 - x^0) D^{(+)}_{\mu \nu} (x, y), \\
S (x, y) &= i \langle 0_e, \mathrm{in} | \mathcal{T} \psi (x) \overline{\psi} (y) | 0_e, \mathrm{in} \rangle \nonumber \\
{}&= \theta (x^0 - y^0) S^{(-)} (x, y) \nonumber \\
{}&- \theta (y^0 - x^0) S^{(+)} (x, y).
\label{eq:S_in_def}
\end{align}
Here $\mathcal{T}$ denotes time ordering. The fermion Green's function $S(x,y)$ is defined with respect to the in-vacuum state; the subscript ``$\mathrm{in}$'' is suppressed only to keep the notation compact. The corresponding in-vacuum expectation value of the current operator is
\begin{align}
j_\mathrm{in}^\mu (x) &= \langle 0_e, \mathrm{in}| j^\mu (x) | 0_e, \mathrm{in} \rangle \nonumber \\
{}&= \frac{ie}{2} \operatorname{Tr} \big \{ \gamma^\mu [S^{(-)} (x, x) - S^{(+)} (x, x)] \big \} \nonumber \\
{}&\equiv ie \operatorname{Tr} \big [\gamma^\mu S(x, x) \big ].
\label{eq:j_in}
\end{align}


\subsection{Generating functional for mean values}

We follow the formalism of Refs.~\cite{fradkin_1981,fradkin_gitman_shvartsman}. The starting point is the generating functional involving two sets of fermionic and photon sources:
\begin{equation}
Z = \langle 0, \mathrm{in} | \mathcal{S}^\dagger (I_2, \eta_2, \overline{\eta}_2) \mathcal{S} (I_1, \eta_1, \overline{\eta}_1) | 0, \mathrm{in} \rangle,
\label{eq:Z_def}
\end{equation}
where
\begin{align}
\mathcal{S} (I, \eta, \overline{\eta}) &= \mathcal{T} \, \mathrm{exp} \, \bigg \{ -i \int \! d^4 x \, \big [ j^\mu (x) A_\mu (x) \nonumber \\
{}&+ I^\mu (x) A_\mu (x) + \overline{\psi} (x) \eta(x) + \overline{\eta} (x) \psi (x) \big ] \bigg \}.
\label{eq:S_source_def}
\end{align}
Here $\eta (x)$ and $\overline{\eta} (x)$ are Grassmann-valued four-component column and row sources, respectively. For zero sources, Eq.~\eqref{eq:S_source_def} reduces to the standard scattering operator $\mathcal{S}$ governing the full real-time evolution in the interaction representation.

As shown in Refs.~\cite{fradkin_1981,fradkin_gitman_shvartsman}, the generating functional~\eqref{eq:Z_def} can be written as
\begin{align}
Z &= \mathrm{exp} \, \bigg \{ e \int \! d^4x' \int \! d^4 y' \int d^4 z' \, \big [\gamma^\mu_{\lambda \alpha \beta} (x', y', z') \big ]_{l l'} \nonumber \\
{}&\times\frac{\delta}{\delta \eta_{\lambda l} (x')} \frac{\delta}{\delta \overline{\eta}_{\alpha l'} (y')} \frac{\delta}{\delta I^\mu_\beta (z')}\bigg \} \, Z_0,
\label{eq:Z_Z0}
\end{align}
where
\begin{align}
\big [\gamma^\mu_{\lambda \alpha \beta} (x', y', z') \big ]_{l l'} &= (\gamma^\mu)_{l l'} \delta_{\lambda \alpha} \delta_{\lambda \beta} \nonumber \\
{}&\times \delta (x'-y') \delta (x' - z') 
\end{align}
and
\begin{align}
Z_0 &= \mathrm{exp} \, \bigg [ i \int \! d^4x \int \! d^4y \, \overline{\eta}_{\alpha l'} (x) S^{\alpha \lambda}_{l' l} (x, y) \eta_{\lambda l} (y) \nonumber \\
{}&- \frac{i}{2} \int \! d^4 z_1 \int d^4 z_2 \, I^\mu_\beta (z_1) D^{\beta \gamma}_{\mu \nu} (z_1, z_2) I^\nu_\gamma (z_2) \bigg ].
\end{align}
The spinor indices $l$ and $l'$ are displayed explicitly, while the indices $\alpha$, $\lambda$, $\beta$, and $\gamma$ take the values $1$ and $2$; summation over repeated indices is understood. These are Keldysh indices and should not be confused with the photon polarization label $\lambda$, which is retained in the standard notation for photon modes. The matrix propagators are
\begin{align}
D^{\beta \gamma}_{\mu \nu} (z_1, z_2) &= \begin{pmatrix}
D_{\mu \nu} (z_1, z_2) & D^{(+)}_{\mu \nu} (z_1, z_2) \\
-D^{(-)}_{\mu \nu} (z_1, z_2) & \overline{D}_{\mu \nu} (z_1, z_2)
\end{pmatrix}_{\beta \gamma}, \label{eq:DS_matrix_def_D} \\
S^{\alpha \lambda} (x, y) &= \begin{pmatrix}
S (x, y) & S^{(+)} (x, y) \\
-S^{(-)} (x, y) & \overline{S} (x, y)
\end{pmatrix}_{\alpha \lambda},
\label{eq:DS_matrix_def_S}
\end{align}
where
\begin{align}
\overline{D}_{\mu \nu} (x, y) &= -i \langle 0_\gamma| \overline{\mathcal{T}} A_\mu (x) A_\nu (y) | 0_\gamma \rangle \nonumber \\
{}&=-\theta (x^0 - y^0) D^{(+)}_{\mu \nu} (x, y) \nonumber \\
{}&+ \theta (y^0 - x^0) D^{(-)}_{\mu \nu} (x, y), \\
\overline{S} (x, y) &= i \langle 0_e, \mathrm{in}| \overline{\mathcal{T}} \psi (x) \overline{\psi} (y) | 0_e, \mathrm{in} \rangle \nonumber \\
{}&= -\theta (x^0 - y^0) S^{(+)} (x, y) \nonumber \\
{}&+ \theta (y^0 - x^0) S^{(-)} (x, y),
\end{align}
and $\overline{\mathcal{T}}$ denotes inverse time ordering. Useful identities obeyed by these functions are collected in Appendix~\ref{sec:app_properties}.

The representation~\eqref{eq:Z_Z0} is convenient for perturbation theory because it contains a single expansion in powers of $e$, in contrast to the original definition~\eqref{eq:Z_def}, which involves a product of two source-dependent scattering operators. To zeroth order, $Z^{(0)} = Z_0$. Once the generating functional is constructed to the desired order, expectation values are obtained by differentiating with respect to the corresponding sources. For example, the mean electromagnetic field is
\begin{equation}
\langle A_\mu (x) \rangle = \frac{i\delta Z}{\delta I^\mu_1 (x)} \bigg |_{I=\eta=\overline{\eta}=0}.
\label{eq:Amean_Z_full}
\end{equation}
In the following sections, we first apply this formalism to the photon number density and then use it to compute the mean electromagnetic field.


\section{Photon number density: nonequilibrium technique} \label{sec:photon_density_ksf}

In this section we compute the mean number density of photons emitted from the initial vacuum state. The calculation is performed within the nonequilibrium technique introduced above, which is particularly convenient here because the external background may violate vacuum stability and the usual in-out formulation is not directly applicable to expectation values. The observable of interest is
\begin{equation}
n_{\mathbf{k} \lambda} = \langle 0,\mathrm{in} | \mathcal{S}^\dagger c^\dagger_{\mathbf{k} \lambda} c_{\mathbf{k} \lambda} \mathcal{S} |0,\mathrm{in}\rangle.
\label{eq:ph_number_density_int}
\end{equation}
Here $\mathbf{k}$ and $\lambda$ denote the momentum and polarization of the emitted photon (physical transverse polarizations correspond to $\lambda=1,2$). We expand Eq.~\eqref{eq:ph_number_density_int} in powers of the electromagnetic coupling, $n_{\mathbf{k}\lambda} = n^{(2)}_{\mathbf{k}\lambda} + n^{(4)}_{\mathbf{k}\lambda} + \ldots$, and evaluate the first two nonvanishing terms. The leading contribution $n^{(2)}_{\mathbf{k}\lambda}$ is of order $e^2$, or $\alpha$, whereas $n^{(4)}_{\mathbf{k}\lambda}$ is of order $e^4$, or $\alpha^2$. The former reproduces the known leading-order result, while the latter is the next-to-leading correction derived below within the same formalism.


\subsection{First order in $\alpha$}

To construct the perturbative expansion of the photon number density~\eqref{eq:ph_number_density_int}, we first introduce the following real-valued function for $x^0$, $y^0 \to +\infty$:
\begin{align}
A_{\mu \nu} (x, y) &= \langle 0,\mathrm{in} | \mathcal{S}^\dagger A_\mu (x) A_\nu (y) \mathcal{S} |0,\mathrm{in}\rangle \nonumber \\
{}&= \frac{\delta^2 Z}{\delta I^\mu_2 (x) \delta I^\nu_1 (y)} \bigg |_{I=\eta=\overline{\eta}=0}.
\end{align}
The leading-order contribution is of order $\alpha$ (second order in $e$) and is evaluated from
\begin{equation}
A^{(2)}_{\mu \nu} (x, y) = \frac{\delta^2 Z^{(2)}}{\delta I^\mu_2 (x) \delta I^\nu_1 (y)} \bigg |_{I=\eta=\overline{\eta}=0}.
\label{eq:Amunu_2_Z}
\end{equation}
Once this function is known, the leading-order number density $n^{(2)}_{\mathbf{k} \lambda}$ is obtained by the projection
\begin{equation}
n^{(2)}_{\mathbf{k} \lambda} = - \int \! d\mathbf{x} \int \! d\mathbf{y} \, f^{\mu}_{\mathbf{k} \lambda} (x) \darrow{\partial}_{x^0} A^{(2)}_{\mu \nu} (x, y) \darrow{\partial}_{y^0} f^{\nu*}_{\mathbf{k} \lambda} (y).
\label{eq:n2_A2}
\end{equation}
To make the calculation transparent, we first compute the leading contribution directly, without referring to diagrams. The explicit form of Eq.~\eqref{eq:Amunu_2_Z} reads
\begin{widetext}
\begin{align}
A^{(2)}_{\mu \nu} (x, y) &= \frac{e^2}{2} \int \! d^4x_1'  \int \! d^4x_2' \, \big (\gamma^{\mu_1} \big )_{l_1 l_1'} (\gamma^{\mu_2} \big )_{l_2 l_2'} \big [ S^{\lambda_1 \lambda_1}_{l_1' l_1} (x_1', x_1') S^{\lambda_2 \lambda_2}_{l_2' l_2} (x_2', x_2') - S^{\lambda_1 \lambda_2}_{l_1' l_2} (x_1', x_2') S^{\lambda_2 \lambda_1}_{l_2' l_1} (x_2', x_1') \big ] \nonumber \\
{}&\times \big [ D^{\lambda_1 \lambda_2}_{\mu_1 \mu_2} (x_1', x_2') D^{12}_{\nu \mu} (y, x) + D^{\lambda_1 2}_{\mu_1 \mu} (x_1', x) D^{\lambda_2 1}_{\mu_2 \nu} (x_2', y) + D^{\lambda_1 1}_{\mu_1 \nu} (x_1', y) D^{\lambda_2 2}_{\mu_2 \mu} (x_2', x) \big ].
\label{eq:Amunu_2_interm}
\end{align}
The first term in the second line does not contribute because of the relations
\begin{equation}
\sum_{\lambda_1, \lambda_2} D^{\lambda_1 \lambda_2}_{\mu_1 \mu_2} (x_1', x_2') = 0, \qquad \sum_{\lambda_1, \lambda_2} D^{\lambda_1 \lambda_2}_{\mu_1 \mu_2} (x_1', x_2') \operatorname{Tr} \big [ \gamma^{\mu_1} S^{\lambda_1 \lambda_2} (x_1', x_2') \gamma^{\mu_2} S^{\lambda_2 \lambda_1} (x_2', x_1') \big ] = 0.
\end{equation}
The first identity removes terms containing a disconnected dumbbell diagram, while the second cancels a disconnected fermionic loop with an internal photon line. This is a general property of the nonequilibrium technique and follows from the fact that each perturbative term $Z^{(k)}$ with $k \geqslant 1$ vanishes at $I = \eta = \overline{\eta} = 0$. Next, the nonzero projections in Eq.~\eqref{eq:n2_A2} arise from the combinations $f_{\mathbf{k} \lambda}^*(x) f_{\mathbf{k} \lambda}(y)$, so it is sufficient to keep only terms with $D^{22} (x_i', x)$ and $D^{11}(x_i', y)$. Finally, using the symmetry of Eq.~\eqref{eq:Amunu_2_interm} under $x_1' \longleftrightarrow x_2'$, we obtain
\begin{align}
A^{(2)}_{\mu \nu} (x, y) &\to e^2 \int \! d^4x_1'  \int \! d^4x_2' \, \big (\gamma^{\mu_1} \big )_{l_1 l_1'} (\gamma^{\mu_2} \big )_{l_2 l_2'} D^{2 2}_{\mu_1 \mu} (x_1', x) D^{1 1}_{\mu_2 \nu} (x_2', y) \nonumber \\
{}&\times \big [ S^{2 2}_{l_1' l_1} (x_1', x_1') S^{1 1}_{l_2' l_2} (x_2', x_2') - S^{2 1}_{l_1' l_2} (x_1', x_2') S^{12}_{l_2' l_1} (x_2', x_1') \big ].
\label{eq:Amunu_2_interm_2}
\end{align}
According to Eq.~\eqref{eq:n2_A2}, this yields
\begin{align}
n^{(2)}_{\mathbf{k} \lambda} &= -e^2 \int \! d^4x_1 \int \! d^4x_2 \, \big (\gamma^{\mu_1} \big )_{l_1 l_1'} (\gamma^{\mu_2} \big )_{l_2 l_2'} f_{\mathbf{k} \lambda \mu_1} (x_1) f^*_{\mathbf{k} \lambda \mu_2} (x_2) \nonumber \\
{}&\times \big [ S^{2 2}_{l_1' l_1} (x_1, x_1) S^{1 1}_{l_2' l_2} (x_2, x_2) - S^{2 1}_{l_1' l_2} (x_1, x_2) S^{12}_{l_2' l_1} (x_2, x_1) \big ].
\label{eq:n2_interm}
\end{align}
\end{widetext}
The first term contains two vacuum currents~\eqref{eq:j_in} and represents the modulus squared of one integral:
\begin{equation}
n^{(2,1)}_{\mathbf{k} \lambda} = \bigg | \int \! d^4x \, j^\mu_\mathrm{in} (x) f^*_{\mathbf{k} \lambda \mu} (x) \bigg |^2.
\label{eq:n2_1}
\end{equation}
The second term is
\begin{align}
n^{(2,2)}_{\mathbf{k} \lambda} &= -e^2 \int \! d^4x_1  \int \! d^4x_2 \, f_{\mathbf{k} \lambda \mu_1} (x_1) f^*_{\mathbf{k} \lambda \mu_2} (x_2) \nonumber \\
{}&\times \operatorname{Tr} \big [ \gamma^{\mu_1} S^{(-)}(x_1, x_2) \gamma^{\mu_2} S^{(+)} (x_2, x_1) \big ].
\label{eq:n2_2}
\end{align}
Using the spectral expansions [see Eqs.~\eqref{eq:Sminus} and \eqref{eq:Splus}], it can be written as
\begin{equation}
n^{(2,2)}_{\mathbf{k} \lambda} = e^2 \sum_{l,s} \bigg | \int \! d^4x \, f^*_{\mathbf{k} \lambda \mu} (x) {}_+ \overline{\varphi}_l (x) \gamma^\mu {}_- \varphi_s (x) \bigg |^2.
\label{eq:n2_2_final}
\end{equation}
Thus the leading-order result can be expressed in terms of standard Feynman diagrams involving the in-solutions and the in-vacuum current~\eqref{eq:j_in}:
\begin{equation}
n^{(2)}_{\mathbf{k} \lambda} =  \bigg | \fdiag[-0.15ex]{1cm}{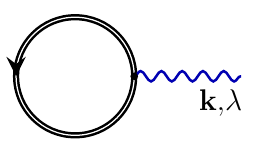} \bigg |^2 + \sum_{l,s} \Bigg | \fdiag[-0.15ex]{1.25cm}{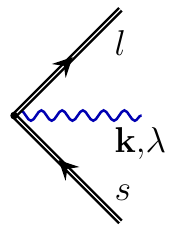} \Bigg |^2.
\label{eq:n_2_final}
\end{equation}
This well-known result was derived, for example, in Refs.~\cite{fradkin_gitman_shvartsman,aleksandrov_prd_2019_2}. As shown in Sec.~\ref{sec:photon_density_direct}, the possibility of rewriting the final expression in terms of usual Feynman diagrams is related to the special structure of the photon number density operator $c^\dagger_{\mathbf{k} \lambda} c_{\mathbf{k} \lambda}$. For a general expectation value, this reduction is no longer available, as will be illustrated in Sec.~\ref{sec:mean_field} for the mean electromagnetic field.

\begin{figure*}[t]
    \centering
    \includegraphics[height=0.1\linewidth]{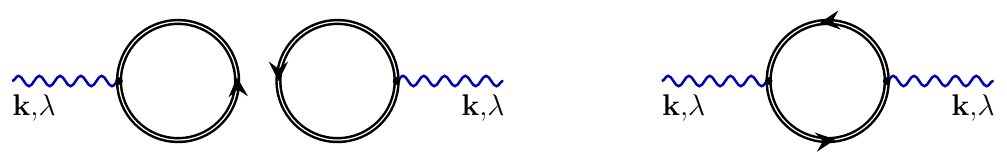}
    \caption{Two diagrams describing the photon number density~\eqref{eq:n2_interm} to first order in $\alpha$ within the nonequilibrium perturbative framework. The incoming and outgoing photon lines correspond to $\lambda_1 = 2$ and $\lambda_2 = 1$, respectively. After the Keldysh-index summation and simplification, they yield Eq.~\eqref{eq:n_2_final}.}
    \label{fig:n2_KSF}
\end{figure*}

The structure of the generating functional~\eqref{eq:Z_Z0} is similar to that of the standard generating functional for the scattering matrix in stable-vacuum QED. Although the present formulation involves matrix propagators, the usual combinatorics and diagrammatic rules still apply, so the relevant channels can be represented by diagrams with matrix-valued propagators. For instance, the leading-order contribution~\eqref{eq:n2_interm} is encoded in the diagrams shown in Fig.~\ref{fig:n2_KSF}. The incoming (outgoing) photon line corresponds to $\lambda=2$ ($\lambda=1$). Vacuum-disconnected loops are automatically removed within the nonequilibrium formulation. We now turn to the next-to-leading contribution of order $e^4$, or $\alpha^2$.

\begin{figure*}[t]
    \centering
    \includegraphics[width=0.8\linewidth]{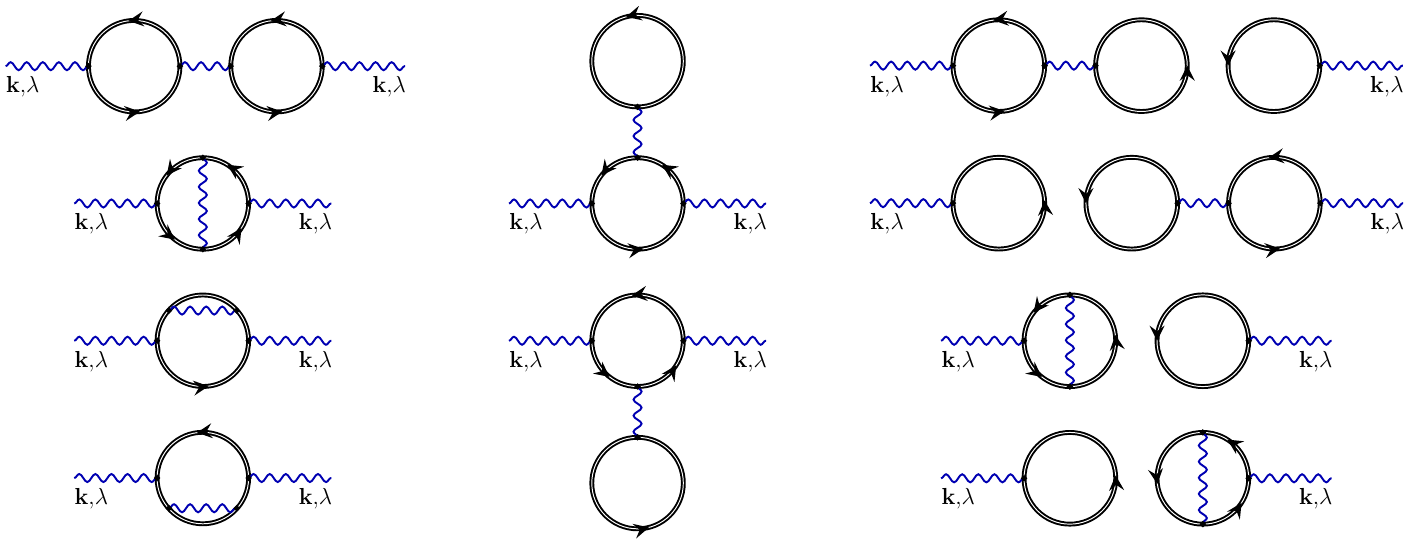}
    \caption{Ten diagrams describing the photon number density~\eqref{eq:n4_ten_terms} to second order in $\alpha$ within the nonequilibrium perturbative framework. After the Keldysh-index summation and simplification, they yield Eq.~\eqref{eq:n_ph_KSF_final}.}
    \label{fig:n4_KSF}
\end{figure*}


\subsection{Second order in $\alpha$}

We now evaluate
\begin{equation}
A^{(4)}_{\mu \nu} (x, y) = \frac{\delta^2 Z^{(4)}}{\delta I^\mu_2 (x) \delta I^\nu_1 (y)} \bigg |_{I=\eta=\overline{\eta}=0}.
\end{equation}
The next-to-leading contribution to the photon number density is then extracted analogously to Eq.~\eqref{eq:n2_A2}:
\begin{equation}
n^{(4)}_{\mathbf{k} \lambda} = - \int \! d\mathbf{x} \int \! d\mathbf{y} \, f^{\mu}_{\mathbf{k} \lambda} (x) \darrow{\partial}_{x^0} A^{(4)}_{\mu \nu} (x, y) \darrow{\partial}_{y^0} f^{\nu*}_{\mathbf{k} \lambda} (y).
\label{eq:n4_A4}
\end{equation}
The resulting expressions contain four integrations over the internal vertices $x_i$ and summation over the internal Keldysh indices $\lambda_3$ and $\lambda_4$, while the external photon lines fix $\lambda_1 = 2$ and $\lambda_2 = 1$. There are 10 nonzero contributions corresponding to the nonequilibrium diagrams displayed in Fig.~\ref{fig:n4_KSF}. Their sum reads
\begin{align}
n^{(4)}_{\mathbf{k} \lambda} &= -ie^4 \int \! d^4x_1 \ldots \int \! d^4x_4 \, (\gamma^{\mu_1})_{l_1 l_1'} \ldots (\gamma^{\mu_4})_{l_4 l_4'} \nonumber \\
{}&\times f_{\mathbf{k} \lambda \mu_1} (x_1) f^*_{\mathbf{k} \lambda \mu_2} (x_2) D_{\mu_3 \mu_4}^{\lambda_3 \lambda_4} (x_3, x_4) \nonumber \\
{}&\times \big ( S_{13} S_{24} S_{31} S_{42} - S_{14} S_{23} S_{31} S_{42} - S_{12} S_{24} S_{31} S_{43} \nonumber \\
{}& - S_{13} S_{21} S_{34} S_{42} + S_{12} S_{23} S_{31} S_{44} + S_{13} S_{21} S_{32} S_{44} \nonumber \\
{}&- S_{13} S_{22} S_{31} S_{44} - S_{11} S_{24} S_{33} S_{42} + S_{14} S_{22} S_{31} S_{43} \nonumber \\
{}& + S_{11} S_{24} S_{32} S_{43} \big ).
\label{eq:n4_ten_terms}
\end{align}
Here we use a compact shorthand notation; for instance, $S_{24} = S^{1\lambda_4}_{l_2' l_4} (x_2, x_4)$. The 10 terms in Eq.~\eqref{eq:n4_ten_terms} can be rewritten in terms of standard Feynman-type diagrams involving the causal Green's functions with respect to the in-vacuum. To this end, one has to perform the summation over $\lambda_3$ and $\lambda_4$. This summation amounts to resolving the possible cuts of the internal propagators and allows one to identify the physical content of the resulting contributions. We illustrate the procedure by considering the last tenth term explicitly:
\begin{align}
n^{(4,10)}_{\mathbf{k} \lambda} &= -ie^4  \int \! d^4x_1 \ldots \int \! d^4x_4 \, (\gamma^{\mu_1})_{l_1 l_1'} \ldots (\gamma^{\mu_4})_{l_4 l_4'} \nonumber \\
{}&\times f_{\mathbf{k} \lambda \mu_1} (x_1) f^*_{\mathbf{k} \lambda \mu_2} (x_2) D_{\mu_3 \mu_4}^{\lambda_3 \lambda_4} (x_3, x_4) \nonumber \\
{}&\times S^{2 2}_{l_1' l_1} (x_1, x_1) S^{1 \lambda_4}_{l_2' l_4} (x_2, x_4) \nonumber \\
{}&\times S^{\lambda_3 1}_{l_3' l_2} (x_3, x_2) S^{\lambda_4 \lambda_3}_{l_4' l_3} (x_4, x_3).
\end{align}
The integral over $x_1$ contains $f_{\mathbf{k} \lambda \mu_1} (x_1)$ contracted with the vacuum current $j^{\mu_1}_\mathrm{in} (x_1)$. The remaining part can be summed directly over $\lambda_3$ and $\lambda_4$, which gives
\begin{widetext}
\begin{align}
n^{(4,10)}_{\mathbf{k} \lambda} &= -e^3 \int \! d^4w \, f_{\mathbf{k} \lambda \sigma} (w) j^{\sigma}_\mathrm{in} (w) \int \! d^4x \int \! d^4y \int \! d^4z \, f^*_{\mathbf{k} \lambda \rho} (z) \Big \{ D_{\mu \nu} (x, y) \operatorname{Tr} \big [ \gamma^\mu S (x,z) \gamma^\rho S (z, y) \gamma^\nu S (y, x) \big ] \nonumber \\
{}&- D^{(+)}_{\mu \nu} (x, y) \operatorname{Tr} \big [ \gamma^\mu S (x,z) \gamma^\rho S^{(+)} (z, y) \gamma^\nu S^{(-)} (y, x) \big ] + D^{(-)}_{\mu \nu} (x, y) \operatorname{Tr} \big [ \gamma^\mu S^{(-)} (x,z) \gamma^\rho S (z, y) \gamma^\nu S^{(+)} (y, x) \big ] \nonumber \\
{}&- \overline{D}_{\mu \nu} (x, y) \operatorname{Tr} \big [ \gamma^\mu S^{(-)} (x,z) \gamma^\rho S^{(+)} (z, y) \gamma^\nu \overline{S} (y, x) \big ]\Big \}.
\end{align}
The first term in braces already has the form of a standard Feynman contribution, whereas the remaining terms can be rewritten using the spectral expansions~\eqref{eq:Sminus} and \eqref{eq:Splus}. It is also useful to express the last term as a complex conjugate. Combining the resulting contributions, we find
\begin{align}
n^{(4,10)}_{\mathbf{k} \lambda} &=-e^3 \int \! d^4w \, f_{\mathbf{k} \lambda \sigma} (w) j^{\sigma}_\mathrm{in} (w) \bigg \{ \int \! d^4x \int \! d^4y \int \! d^4z \, f^*_{\mathbf{k} \lambda \rho} (z) D_{\mu \nu} (x, y) \operatorname{Tr} \big ( \gamma^\mu S \gamma^\rho S \gamma^\nu S \big ) \nonumber \\
{}&+i \sum_{l,s} \sum_{\mathbf{k}', \lambda'} \int \! d^4x \int \! d^4y \int \! d^4z \, f^*_{\mathbf{k} \lambda \rho} (z) f^*_{\mathbf{k}' \lambda' \mu} (x) f_{\mathbf{k}' \lambda' \nu} (y) \big ( {}_+ \overline{\varphi}_l \gamma^\mu S \gamma^\rho {}_- \varphi_s \big ) \big ( {}_- \overline{\varphi}_s \gamma^\nu {}_+ \varphi_l \big ) \nonumber \\
{}&+i \sum_{l,s} \sum_{\mathbf{k}', \lambda'} \int \! d^4x \int \! d^4y \int \! d^4z \, f^*_{\mathbf{k} \lambda \rho} (z) f_{\mathbf{k}' \lambda' \mu} (x) f^*_{\mathbf{k}' \lambda' \nu} (y) \big ( {}_- \overline{\varphi}_s \gamma^\mu {}_+ \varphi_l \big ) \big ( {}_+ \overline{\varphi}_l \gamma^\rho S \gamma^\nu {}_- \varphi_s \big )  \nonumber \\
{}&+ \sum_{l,s} \int \! d^4x \int \! d^4y \int \! d^4z \, \Big [ f_{\mathbf{k} \lambda \rho} (z) D_{\mu \nu} (x,y) \big ( {}_+ \overline{\varphi}_l \gamma^\mu S \gamma^\nu {}_- \varphi_s \big ) \big ( {}_- \overline{\varphi}_s \gamma^\rho {}_+ \varphi_l \big ) \Big ]^* \bigg \},
\end{align}
where the arguments of the fermionic functions have been suppressed. This expression can be represented compactly in terms of standard Feynman diagrams:
\begin{align}
n^{(4,10)}_{\mathbf{k} \lambda} &= \bigg ( \fdiag[-0.15ex]{1cm}{tadpole.pdf} \bigg )^{\!\!*} \Bigg \{ \fdiag[-0.15ex]{1cm}{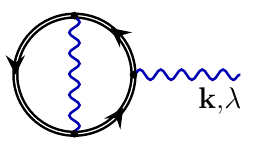} + \sum_{l,s} \sum_{\mathbf{k}', \lambda'} \Bigg ( \fdiag[-0.15ex]{1.3cm}{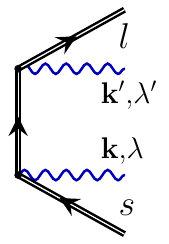} \Bigg ) \Bigg ( \fdiag[-0.15ex]{1.25cm}{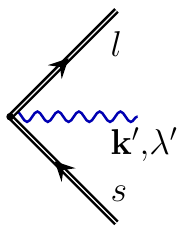} \Bigg )^{\!\!*} \nonumber \\
{}& + \sum_{l,s} \sum_{\mathbf{k}', \lambda'} \Bigg ( \fdiag[-0.15ex]{1.3cm}{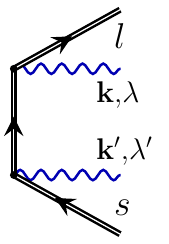} \Bigg ) \Bigg ( \fdiag[-0.15ex]{1.25cm}{vertex_kp_ls.pdf} \Bigg )^{\!\!*} + \sum_{l,s} \Bigg ( \fdiag[-0.15ex]{1.25cm}{vertex.pdf} \Bigg ) \Bigg ( \fdiag[-0.15ex]{1.3cm}{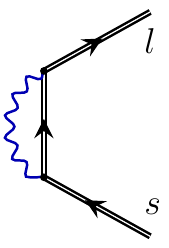} \, \Bigg )^{\!\!*}\Bigg \}.
\end{align}
Here the fermionic lines correspond either to one-particle in-solutions or to causal Green's functions with respect to the in-vacuum. Performing analogous calculations for the remaining nine terms in Eq.~\eqref{eq:n4_ten_terms}, we obtain the final result:
\begin{align}
n^{(4)}_{\mathbf{k} \lambda} &= \sum_{\mathbf{k}', \lambda'} \bigg | \fdiag[-0.15ex]{1cm}{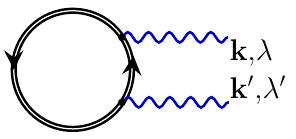} \bigg |^2 + \sum_{l,s} \sum_{\mathbf{k}', \lambda'} \Bigg | \fdiag[-0.15ex]{1.3cm}{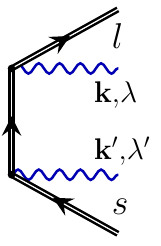} + \fdiag[-0.15ex]{1.3cm}{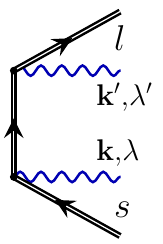} \Bigg |^2 \nonumber  \\
{}&+ 2 \operatorname{Re} \Bigg [ \bigg ( \fdiag[-0.15ex]{1cm}{tadpole.pdf} \bigg )^{\!\! *} \bigg ( \fdiag[-0.15ex]{1cm}{tadpole_int_photon.pdf} + \fdiag[-0.15ex]{1cm}{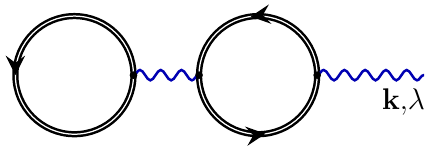} \bigg ) \Bigg ] \nonumber \\
{}&+ 2 \operatorname{Re} \sum_{\mathbf{k}', \lambda'} \, \Bigg [ \bigg ( \fdiag[-0.15ex]{1cm}{tadpole.pdf}~\fdiag[-0.15ex]{1cm}{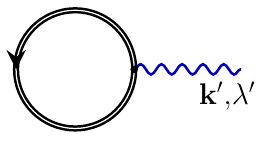} \bigg )^{\!\! *}~~\fdiag[-0.15ex]{1cm}{tadpole_double.pdf} \Bigg ] \nonumber \\
{}&+ 2 \operatorname{Re} \sum_{l,s} \, \Bigg \{ \Bigg ( \fdiag[-0.15ex]{1.25cm}{vertex.pdf} \Bigg )^{\!\! *} \Bigg [ \Bigg ( \fdiag[-0.15ex]{1.3cm}{pair_loop.pdf}~\fdiag[-0.15ex]{1.0cm}{tadpole.pdf} \Bigg ) + \Bigg ( \fdiag[-0.15ex]{0.9cm}{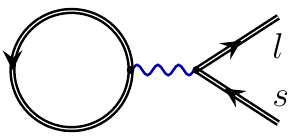}~\fdiag[-0.15ex]{1.0cm}{tadpole.pdf} \Bigg ) \nonumber \\
{}& + \fdiag[-0.15ex]{1.0cm}{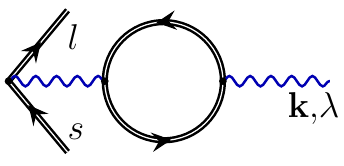} + \fdiag[-0.15ex]{1.3cm}{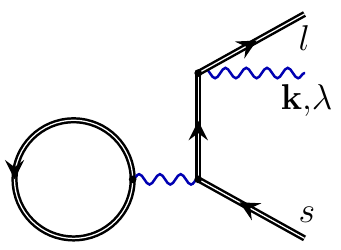} + \fdiag[-0.15ex]{1.3cm}{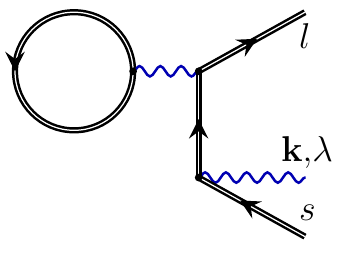} + \, \fdiag[-0.15ex]{1.3cm}{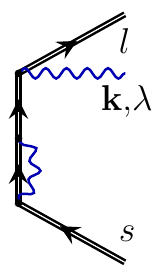} + \, \fdiag[-0.15ex]{1.3cm}{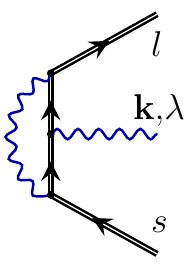} + \, \fdiag[-0.15ex]{1.3cm}{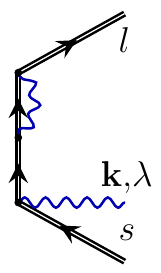}\Bigg ] \Bigg \} \nonumber \\
{}&+ 2 \operatorname{Re} \sum_{l,s} \sum_{\mathbf{k}', \lambda'} \Bigg [ \Bigg ( \fdiag[-0.15ex]{1.25cm}{vertex.pdf}~\fdiag[-0.15ex]{1.0cm}{tadpole_kp.pdf} \Bigg )^{\!\! *} \Bigg ( \fdiag[-0.15ex]{1.25cm}{vertex_kp_ls.pdf}~\fdiag[-0.15ex]{1.0cm}{tadpole.pdf} \Bigg ) \Bigg ] \nonumber \\
{}&+ 2 \operatorname{Re} \sum_{l,s} \sum_{\mathbf{k}', \lambda'} \Bigg [ \Bigg ( \fdiag[-0.15ex]{1.25cm}{vertex.pdf}~\fdiag[-0.15ex]{1.0cm}{tadpole_kp.pdf} ~+~ \fdiag[-0.15ex]{1.25cm}{vertex_kp_ls.pdf}~\fdiag[-0.15ex]{1.0cm}{tadpole.pdf} \Bigg )^{\!\! *} \Bigg ( \fdiag[-0.15ex]{1.3cm}{pair_two_photons_1.pdf} ~+~ \fdiag[-0.15ex]{1.3cm}{pair_two_photons_2.pdf} \Bigg ) \Bigg ] \nonumber \\
{}& + \operatorname{Re} \sum_{l,s,r,t} \sum_{\mathbf{k}', \lambda'} \Bigg \{ \Bigg ( \fdiag[-0.15ex]{1.25cm}{vertex.pdf}~\fdiag[-0.15ex]{1.25cm}{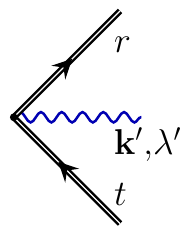}\Bigg )^{\!\! *} \Bigg [ \Bigg ( \fdiag[-0.15ex]{1.25cm}{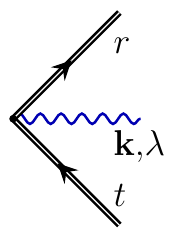}~\fdiag[-0.15ex]{1.25cm}{vertex_kp_ls.pdf}\Bigg ) - \Bigg ( \fdiag[-0.15ex]{1.25cm}{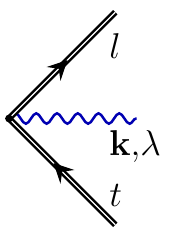}~\fdiag[-0.15ex]{1.25cm}{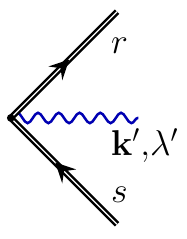}\Bigg ) - \Bigg ( \fdiag[-0.15ex]{1.25cm}{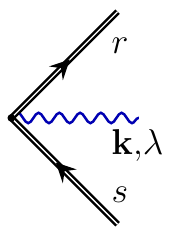}~\fdiag[-0.15ex]{1.25cm}{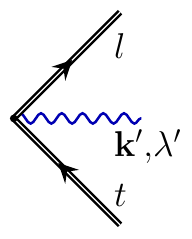}\Bigg )\Bigg ] \Bigg \}.
\label{eq:n_ph_KSF_final}
\end{align}
\end{widetext}
The terms in Eq.~\eqref{eq:n_ph_KSF_final} are arranged according to their physical origin. The first line contains the pure two-photon channels: radiation by the induced current and pair creation accompanied by two photons. The next two lines describe the interference of the leading tadpole amplitude with higher-order induced-current contributions, including the channel with an additional unobserved photon. The following terms collect corrections to, and interferences with, the leading vertex and tadpole mechanisms. The last line contains the purely fermionic exchange and interference structures generated by different pair-production amplitudes. In this form the long expression separates the distinct radiative mechanisms while keeping the result compact.

The loop contributions entering Eq.~\eqref{eq:n_ph_KSF_final} and the mean-field result below are written in a formal regularized sense. Their ultraviolet structure is the same as in QED in external backgrounds and requires the usual renormalization of the electromagnetic field, charge, and local vacuum-polarization terms. A detailed renormalization analysis is not needed for the diagrammatic organization developed here and will depend on the concrete external-field configuration used in applications.


\section{Photon number density: direct evaluation} \label{sec:photon_density_direct}

Although all contributions have already been identified using the Keldysh-Schwinger-Fradkin nonequilibrium technique, it is instructive to derive $n^{(4)}_{\mathbf{k} \lambda}$ independently from the identity
\begin{align}
n_{\mathbf{k} \lambda} &= \langle 0,\mathrm{in} | \mathcal{S}^\dagger c^\dagger_{\mathbf{k} \lambda} c_{\mathbf{k} \lambda} \mathcal{S} |0,\mathrm{in}\rangle \nonumber \\
{}&=\sum_{|\mathrm{in}\rangle} \big | \langle \mathrm{in} | c_{\mathbf{k} \lambda} \mathcal{S} |0,\mathrm{in}\rangle \big |^2.
\label{eq:n_in}
\end{align}
This approach not only confirms the result obtained above, but also clarifies the structure of Eq.~\eqref{eq:n_ph_KSF_final}. Since we keep terms only through order $\alpha^2$, the operator $\mathcal{S}$ in Eq.~\eqref{eq:n_in} can be replaced by its first few perturbative contributions. Accordingly, only a finite number of in-states contribute to $n^{(2)}_{\mathbf{k} \lambda}$ and $n^{(4)}_{\mathbf{k} \lambda}$, and these states can be analyzed separately. This also explains why the final result~\eqref{eq:n_ph_KSF_final} can be written in terms of standard Feynman-diagram ingredients: each amplitude entering Eq.~\eqref{eq:n_in} is an ordinary $S$-matrix element between in-states, while the interference terms arise from taking the modulus squared.

The general structure of the perturbative expansion of $n_{\mathbf{k} \lambda}$ has the following form:
\begin{widetext}
\begin{align}
n_{\mathbf{k} \lambda} &= \bigg | \fdiag[-0.15ex]{1cm}{tadpole.pdf} + \bigg ( \fdiag[-0.15ex]{1cm}{tadpole.pdf}~\fdiag[-0.15ex]{1cm}{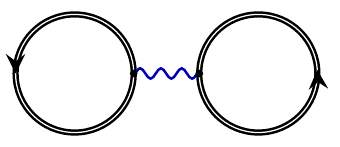} \bigg ) + \bigg ( \fdiag[-0.15ex]{1cm}{tadpole.pdf}~\fdiag[-0.20ex]{1cm}{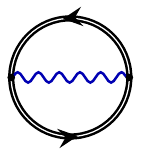} \bigg ) \nonumber \\
{}&+ \fdiag[-0.15ex]{1cm}{tadpole_int_photon.pdf} + \fdiag[-0.15ex]{1cm}{tadpole_two.pdf} + \mathcal{O} (e^5) \bigg |^2 \nonumber \\
{}&+\sum_{\mathbf{k}', \lambda'} \bigg | \fdiag[-0.15ex]{1cm}{tadpole_double.pdf} + \bigg ( \fdiag[-0.15ex]{1cm}{tadpole.pdf}~\fdiag[-0.15ex]{1cm}{tadpole_kp.pdf} \bigg ) + \mathcal{O} (e^4) \bigg |^2 \nonumber \\
{}&+\sum_{l,s} \Bigg | \fdiag[-0.15ex]{1.25cm}{vertex.pdf} + \Bigg ( \fdiag[-0.15ex]{1.25cm}{vertex.pdf}~\fdiag[-0.15ex]{1cm}{vacuum_dumbbell.pdf} \Bigg ) + \Bigg ( \fdiag[-0.15ex]{1.25cm}{vertex.pdf}~\fdiag[-0.20ex]{1cm}{vacuum_theta.pdf} \Bigg ) + \Bigg ( \fdiag[-0.15ex]{0.9cm}{tadpole_pair.pdf}~\fdiag[-0.15ex]{1.0cm}{tadpole.pdf} \Bigg ) \nonumber \\
{}&+ \fdiag[-0.15ex]{1.0cm}{pair_loop_photon.pdf} + \Bigg ( \fdiag[-0.15ex]{1.3cm}{pair_loop.pdf}~\fdiag[-0.15ex]{1.0cm}{tadpole.pdf} \Bigg ) + \fdiag[-0.15ex]{1.3cm}{pair_photon_tadpole_1.pdf} + \fdiag[-0.15ex]{1.3cm}{pair_photon_tadpole_2.pdf} + \, \fdiag[-0.15ex]{1.3cm}{pair_photon_loop_1.pdf} + \, \fdiag[-0.15ex]{1.3cm}{vertex_loop.pdf} + \, \fdiag[-0.15ex]{1.3cm}{pair_photon_loop_2.pdf} + \mathcal{O} (e^5) \Bigg |^2 \nonumber \\
{}&+ \sum_{l,s} \sum_{\mathbf{k}', \lambda'} \Bigg | \fdiag[-0.15ex]{1.3cm}{pair_two_photons_1.pdf} ~+~ \fdiag[-0.15ex]{1.3cm}{pair_two_photons_2.pdf} + \Bigg ( \fdiag[-0.15ex]{1.25cm}{vertex.pdf}~\fdiag[-0.15ex]{1.0cm}{tadpole_kp.pdf}\Bigg ) + \Bigg ( \fdiag[-0.15ex]{1.25cm}{vertex_kp_ls.pdf}~\fdiag[-0.15ex]{1.0cm}{tadpole.pdf} \Bigg ) + \mathcal{O} (e^4) \Bigg |^2 \nonumber  \\
{}&+ \frac{1}{4} \sum_{l,s,r,t} \sum_{\mathbf{k}', \lambda'} \Bigg | \Bigg ( \fdiag[-0.15ex]{1.25cm}{vertex.pdf}~\fdiag[-0.15ex]{1.25cm}{vertex_kp_rt.pdf}\Bigg ) - \Bigg ( \fdiag[-0.15ex]{1.25cm}{vertex_rs.pdf}~\fdiag[-0.15ex]{1.25cm}{vertex_kp_lt.pdf}\Bigg ) + \Bigg ( \fdiag[-0.15ex]{1.25cm}{vertex_rt.pdf}~\fdiag[-0.15ex]{1.25cm}{vertex_kp_ls.pdf}\Bigg ) - \Bigg ( \fdiag[-0.15ex]{1.25cm}{vertex_lt.pdf}~\fdiag[-0.15ex]{1.25cm}{vertex_kp_rs.pdf}\Bigg ) + \mathcal{O}(e^4) \Bigg |^2 + \mathcal{O}(e^6).
\label{eq:n_ph_direct}
\end{align}
\end{widetext}
The representation~\eqref{eq:n_ph_direct} contains both contributions to $n^{(2)}_{\mathbf{k} \lambda}$ and makes it possible to trace all terms entering $n^{(4)}_{\mathbf{k} \lambda}$, including interference contributions. At the same time, this direct amplitude-based form contains vacuum-disconnected factors. They appear because the transition amplitudes in Eq.~\eqref{eq:n_in} are ordinary $S$-matrix elements between in-states and therefore include the vacuum persistence amplitude as a multiplicative factor. As discussed in Sec.~\ref{sec:photon_density_ksf}, vacuum-disconnected contributions are automatically removed in the nonequilibrium calculation. In the direct approach, their cancellation is not built in and must be performed explicitly.

The required cancellation is provided by the optical-theorem identities appropriate for the unstable-vacuum setup:
\begin{align}
2 \operatorname{Im} \fdiag[-0.15ex]{1cm}{vacuum_dumbbell.pdf} &= \sum_{\mathbf{k}', \lambda'} \bigg | \fdiag[-0.15ex]{1cm}{tadpole_kp.pdf} \bigg |^2, \label{eq:OT_1} \\
2 \operatorname{Im} \fdiag[-0.20ex]{1cm}{vacuum_theta.pdf} &= \sum_{l,s} \sum_{\mathbf{k}', \lambda'} \Bigg | \fdiag[-0.15ex]{1.25cm}{vertex_kp_ls.pdf} \Bigg |^2.
\label{eq:OT_2}
\end{align}
These relations follow from the unitarity condition for the $S$-operator in the unstable-vacuum setup; an analogous optical-theorem relation also holds, e.g., for the polarization tensor~\cite{aleksandrov_ois_2025}. The optical theorem is the precise mechanism by which the vacuum-disconnected pieces in Eq.~\eqref{eq:n_ph_direct} are canceled by the corresponding terms involving final-state summations. After applying Eqs.~\eqref{eq:OT_1} and \eqref{eq:OT_2} and retaining terms through order $\alpha^2$, the direct calculation reproduces the result~\eqref{eq:n_ph_KSF_final}. This agreement is a nontrivial check of the nonequilibrium derivation and clarifies why the photon number density, unlike a generic expectation value, admits an equivalent amplitude-level representation.


\section{Mean electromagnetic field} \label{sec:mean_field}

In this section, instead of the photon number density operator $c^\dagger_{\mathbf{k} \lambda} c_{\mathbf{k} \lambda}$, we consider the mean electromagnetic field~\eqref{eq:Amean_Z_full}. The leading contribution, of order $e$, is given by
\begin{align}
\langle A_\mu (x) \rangle^{(1)} &= \frac{i\delta Z^{(1)}}{\delta I^\mu_1 (x)} \bigg |_{I=\eta=\overline{\eta}=0} \nonumber \\
{}&= \int \! d^4 x' \, j^\nu_\mathrm{in} (x') \big [ D^{1 1}_{\nu \mu} (x', x) + D^{2 1}_{\nu \mu} (x', x) \big ] \nonumber \\
{}&= \int \! d^4 x' \, j^\nu_\mathrm{in} (x') \big [ D_{\mu \nu} (x, x') + D^{(+)}_{\mu \nu} (x, x') \big ] \nonumber \\
{}&= \int \! d^4 x' \, D^{(\mathrm{R})}_{\mu \nu} (x, x') j^\nu_\mathrm{in} (x').
\label{eq:Amean_1_final}
\end{align}
This result was discussed, e.g., in Ref.~\cite{fradkin_gitman_shvartsman}. The leading-order mean field is generated by the source induced by the external background. The retarded photon Green's function in Eq.~\eqref{eq:Amean_1_final} ensures causality; the same retarded structure follows from the all-order Schwinger-Dyson equation discussed in Appendix~\ref{sec:app_schwinger-dyson}. The relations between the retarded and advanced Green's functions and the photon functions introduced above are summarized in Appendix~\ref{sec:app_properties}. We emphasize that the in-vacuum current $j^\mu_{\mathrm{in}}$ contains both vacuum-polarization effects and contributions associated with real particles produced from the vacuum, although this separation is not unambiguously defined at intermediate times (see, e.g., Ref.~\cite{aleksandrov_arxiv_2026}).

\begin{figure*}[t]
    \centering
    \includegraphics[height=0.1\linewidth]{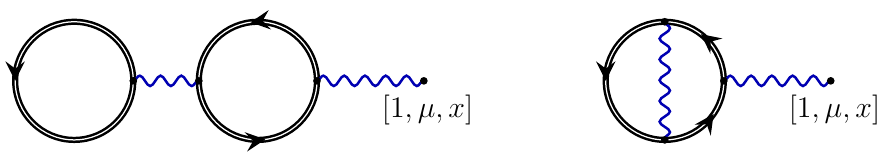}
    \caption{Two diagrams describing the mean electromagnetic field $\langle A_\mu (x)\rangle$ to third order in $e$ within the nonequilibrium perturbative framework.}
    \label{fig:A3_KSF}
\end{figure*}

We now calculate the third-order ($e^3$) contribution,
\begin{equation}
\langle A_\mu (x) \rangle^{(3)} = \frac{i\delta Z^{(3)}}{\delta I^\mu_1 (x)} \bigg |_{I=\eta=\overline{\eta}=0}.
\end{equation}
There are two diagrams contributing to this quantity, shown in Fig.~\ref{fig:A3_KSF}. The explicit form of the third-order correction reads
\begin{widetext}
\begin{align}
\langle A_\sigma (w) \rangle^{(3)} &= - \int \! d^4x \int \! d^4y \int \! d^4z \, D^{\lambda_1 \lambda_2}_{\mu \nu} (x, y) D^{\lambda_3 1}_{\rho \sigma} (z, w) \nonumber \\
{}&\times \Big \{ ie^2 j^{\mu}_\mathrm{in} (x) \operatorname{Tr} \big [ \gamma^\nu S^{\lambda_2 \lambda_3} (y, z) \gamma^\rho S^{\lambda_3 \lambda_2} (z, y) \big ] + e^3 \operatorname{Tr} \big [ \gamma^\mu S^{\lambda_1 \lambda_2} (x, y) \gamma^\nu S^{\lambda_2 \lambda_3} (y, z) \gamma^\rho S^{\lambda_3 \lambda_1} (z, x) \big ] \Big \}.
\label{eq:A3}
\end{align}
\end{widetext}
We evaluate the two terms separately. First we isolate the part that could make the first contribution complex. Using Eqs.~\eqref{eq:Dconj_gen} and \eqref{eq:Sconj_gen}, the difference between this term and its complex conjugate is proportional to
\begin{equation}
D^{\lambda_1 \lambda_2}_{\mu \nu} (x, y) D^{\lambda_3 \lambda_4}_{\rho \sigma} (z, w) \operatorname{Tr} \big [ \gamma^\nu S^{\lambda_2 \lambda_3} (y, z) \gamma^\rho S^{\lambda_3 \lambda_2} (z, y) \big ],
\label{eq:DDTr_zero}
\end{equation}
where summation over all $\lambda_i$ is implied. The expression vanishes because
\begin{equation}
(-1)^{\lambda_2 + \lambda_3} \operatorname{Tr} \big [ \gamma^\nu S^{\lambda_2 \lambda_3} (y, z) \gamma^\rho S^{\lambda_3 \lambda_2} (z, y) \big ] = 0,
\end{equation}
as follows by considering separately the domains $y^0 > z^0$ and $y^0 < z^0$. Thus the expression~\eqref{eq:DDTr_zero} is zero; diagrammatically, this is the cancellation of the disconnected vacuum graph with three fermionic bubbles connected in series by two photon lines. The same reasoning applies to the second contribution. In that case one encounters
\begin{multline*}
D^{\lambda_1 \lambda_2}_{\mu \nu} (x, y) D^{\lambda_3 \lambda_4}_{\rho \sigma} (z, w) \\
{}\times \operatorname{Tr} \big [ \gamma^\mu S^{\lambda_1 \lambda_2} (x, y) \gamma^\nu S^{\lambda_2 \lambda_3} (y, z) \gamma^\rho S^{\lambda_3 \lambda_1} (z, x) \big ] = 0.
\end{multline*}
which corresponds to the vanishing of the disconnected vacuum diagram containing a dumbbell with an additional internal photon line in the fermionic loop.

Since both terms in Eq.~\eqref{eq:A3} are real, we may keep only the real parts after summing over the indices $\lambda_i$. For the contribution generated by the leading induced current $j^\mu_\mathrm{in}$, we use
\begin{align}
\sum_{\lambda_1} D^{\lambda_1 \lambda_2}_{\mu \nu} (x, y) &= (-1)^{\lambda_2 + 1} D^{(\mathrm{A})}_{\mu \nu} (x, y) \nonumber \\
{}&= (-1)^{\lambda_2 + 1} D^{(\mathrm{R})}_{\nu \mu} (y, x)
\end{align}
and then sum over $\lambda_2$ and $\lambda_3$:
\begin{align}
\langle A_\sigma (w) \rangle^{(3)}_1 &= e^2 \int \! d^4x \int \! d^4y \int \! d^4z \, j^{\mu}_\mathrm{in} (x) D^{(\mathrm{R})}_{\nu \mu} (y, x) \nonumber \\
{}&\times \operatorname{Im} \Big \{ D_{\sigma \rho} (w, z) \operatorname{Tr} \big [ \gamma^\nu S (y, z) \gamma^\rho S (z, y) \big ] \nonumber \\
{}&+ D_{\sigma \rho} (w, z) \operatorname{Tr} \big [ \gamma^\nu S^{(-)} (y, z) \gamma^\rho S^{(+)} (z, y) \big ] \nonumber \\
{}&- D^{(+)}_{\sigma \rho} (w, z) \operatorname{Tr} \big [ \gamma^\nu S^{(+)} (y, z) \gamma^\rho S^{(-)} (z, y) \big ] \nonumber \\
{}&- D^{(+)}_{\sigma \rho} (w, z) \operatorname{Tr} \big [ \gamma^\nu \overline{S} (y, z) \gamma^\rho \overline{S} (z, y) \big ] \Big \}.
\end{align}
It is convenient to replace the last two terms in braces with their complex conjugates taken with opposite signs. Then
\begin{align}
\operatorname{Im} \big \{ \ldots \big \} &= \operatorname{Im}  \Big ( \big [ D_{\sigma \rho} (w, z) + D^{(-)}_{\sigma \rho} (w, z) \big ] \nonumber \\
{}&\times \big \{ \operatorname{Tr} \big [ \gamma^\nu S (y, z) \gamma^\rho S (z, y) \big ] \nonumber \\
{}&+ \operatorname{Tr} \big [ \gamma^\nu S^{(-)} (y, z) \gamma^\rho S^{(+)} (z, y) \big ] \big \} \Big ).
\label{eq:A13_im}
\end{align}
The expression in braces vanishes for $y^0 > z^0$ and can be rewritten as
\begin{align}
&\operatorname{Tr} \big [ \gamma^\nu S (y, z) \gamma^\rho S (z, y) \big ] + \operatorname{Tr} \big [ \gamma^\nu S^{(-)} (y, z) \gamma^\rho S^{(+)} (z, y) \big ] \nonumber \\
{}&= 2 i \theta (z^0 - y^0) \operatorname{Im} \operatorname{Tr} \big [ \gamma^\nu S^{(-)} (y, z) \gamma^\rho S^{(+)} (z, y) \big ] \nonumber \\
{}&= 2 i \theta (z^0 - y^0) \operatorname{Im} \operatorname{Tr} \big [ \gamma^\nu S (y, z) \gamma^\rho S (z, y) \big ].
\end{align}
This factor is purely imaginary, so in Eq.~\eqref{eq:A13_im} one has to take
\begin{equation}
\operatorname{Re} \big [ D_{\sigma \rho} (w, z) + D^{(-)}_{\sigma \rho} (w, z) \big ] = D^{(\mathrm{R})}_{\sigma \rho} (w, z).
\end{equation}
Combining these results, we obtain
\begin{align}
\langle A_\sigma (w) \rangle^{(3)}_1 &= 2 \int \! d^4x \int \! d^4y \int \! d^4z \, D^{(\mathrm{R})}_{\sigma \mu} (w, x) \theta (x^0 - y^0) \nonumber \\
{}&\times \big [ \operatorname{Re} \pi^{\mu \nu} (x,y) \big ] D^{(\mathrm{R})}_{\nu \rho} (y, z) j^{\rho}_\mathrm{in} (z),
\label{eq:A3_1_final}
\end{align}
where
\begin{equation}
\pi^{\mu \nu} (x,y) = -ie^2 \operatorname{Tr} \big [ \gamma^\mu S (x, y) \gamma^\nu S (y, x) \big ].
\end{equation}

We now turn to the second term in Eq.~\eqref{eq:A3}:
\begin{widetext}
\begin{equation}
\langle A_\sigma (w) \rangle^{(3)}_2 = -e^3 \int \! d^4x \int \! d^4y \int \! d^4z \, D^{\lambda_1 \lambda_2}_{\mu \nu} (x, y) D^{\lambda_3 1}_{\rho \sigma} (z, w) \operatorname{Tr} \big [ \gamma^\mu S^{\lambda_1 \lambda_2} (x, y) \gamma^\nu S^{\lambda_2 \lambda_3} (y, z) \gamma^\rho S^{\lambda_3 \lambda_1} (z, x) \big ].
\label{eq:A3_2_initial}
\end{equation}
Here the indices $\lambda_1$, $\lambda_2$, and $\lambda_3$ have to be summed over. The Keldysh-index sum is most transparent when the integration domain is split according to the relative order of $x^0$, $y^0$, and $z^0$. All regions cancel except those in which $z^0$ is later than both $x^0$ and $y^0$. The two remaining domains, $x^0 > y^0$ and $y^0 > x^0$, give
\begin{align}
\langle A_\sigma (w) \rangle^{(3)}_2 &= e^3 \int \! d^4x \int \! d^4y \int \! d^4z \, D^{(\mathrm{R})}_{\sigma \rho} (w, z) \theta (z^0 - x^0) \theta (z^0 - y^0) \nonumber \\
{}&\times \Big ( \theta (x^0 - y^0) \Big \{ D^{(+)}_{\mu \nu} (x,y) \operatorname{Tr} \big [ \gamma^\mu S^{(+)} (x,y) \gamma^\nu S^{(-)} (y,z) \gamma^\rho S^{(0)} (z,x) \big ] \nonumber \\
{}&+ D^{(-)}_{\mu \nu} (x,y) \operatorname{Tr} \big [ \gamma^\mu S^{(-)} (x,y) \gamma^\nu S^{(+)} (y,z) \gamma^\rho S^{(0)} (z,x) \big ] \Big \} \nonumber \\
{}&+\theta (y^0 - x^0) \Big \{ D^{(+)}_{\mu \nu} (x,y) \operatorname{Tr} \big [ \gamma^\mu S^{(+)} (x,y) \gamma^\nu S^{(0)} (y,z) \gamma^\rho S^{(-)} (z,x) \big ] \nonumber \\
{}&+ D^{(-)}_{\mu \nu} (x,y) \operatorname{Tr} \big [ \gamma^\mu S^{(-)} (x,y) \gamma^\nu S^{(0)} (y,z) \gamma^\rho S^{(+)} (z,x) \big ] \Big \} \Big ).
\label{eq:A3_2_interm}
\end{align}
In analogy with the photon spectral function $D^{(0)}_{\mu\nu}$, we have introduced
\begin{equation}
S^{(0)}(x,y) = S^{(+)}(x,y) + S^{(-)}(x,y) = i \big \{\psi(x),\overline{\psi}(y) \big \}.
\label{eq:S0}
\end{equation}
This object should not be confused with a free fermion propagator; it is the spectral, or Pauli-Jordan, kernel of the Dirac field in the external background. Since the anticommutator in Eq.~\eqref{eq:S0} is a $c$-number, this kernel coincides with its in- or out-vacuum expectation value.

By performing complex conjugation and then interchanging $x \longleftrightarrow y$ and $\mu \longleftrightarrow \nu$ in the term containing $\theta (x^0 - y^0)$ in Eq.~\eqref{eq:A3_2_interm}, one finds that this contribution is the complex conjugate of the term containing $\theta (y^0 - x^0)$. The result can therefore be written as
\begin{align}
\langle A_\sigma (w) \rangle^{(3)}_2 &= 2e^3 \int \! d^4x \int \! d^4y \int \! d^4z \, D^{(\mathrm{R})}_{\sigma \rho} (w, z) \theta (z^0 - y^0) \theta (y^0 - x^0) \nonumber \\
{}&\times \operatorname{Re} \Big \{ D^{(+)}_{\mu \nu} (x,y) \operatorname{Tr} \big [ \gamma^\mu S^{(+)} (x,y) \gamma^\nu S^{(0)} (y,z) \gamma^\rho S^{(-)} (z,x) \big ] \nonumber \\
{}&+ D^{(-)}_{\mu \nu} (x,y) \operatorname{Tr} \big [ \gamma^\mu S^{(-)} (x,y) \gamma^\nu S^{(0)} (y,z) \gamma^\rho S^{(+)} (z,x) \big ] \Big \}.
\label{eq:A3_2_final}
\end{align}
Combining Eqs.~\eqref{eq:A3_1_final} and \eqref{eq:A3_2_final}, we obtain the full third-order contribution to the mean field:
\begin{align}
\langle A_\sigma (w) \rangle^{(3)} &= 2 \int \! d^4x \int \! d^4y \int \! d^4z \, D^{(\mathrm{R})}_{\sigma \mu} (w, x) \theta (x^0 - y^0) \Big ( \big [ \operatorname{Re} \pi^{\mu \nu} (x,y) \big ] D^{(\mathrm{R})}_{\nu \rho} (y, z) j^{\rho}_\mathrm{in} (z) \nonumber \\
{}&+ e^3 \theta (y^0 - z^0) \operatorname{Re} \Big \{ D^{(+)}_{\rho \nu} (z,y) \operatorname{Tr} \big [ \gamma^\mu S^{(-)} (x,z) \gamma^\rho S^{(+)} (z,y) \gamma^\nu S^{(0)} (y,x) \big ] \nonumber \\
{}&+ D^{(-)}_{\rho \nu} (z,y) \operatorname{Tr} \big [ \gamma^\mu S^{(+)} (x,z) \gamma^\rho S^{(-)} (z,y) \gamma^\nu S^{(0)} (y,x) \big ] \Big \} \Big ).
\label{eq:A3_full_final}
\end{align}
\end{widetext}
Equation~\eqref{eq:A3_full_final} gives an explicit prescription for computing the unrenormalized third-order correction to the mean electromagnetic field in a general spacetime-dependent external background. Its causal structure is manifest: the field observed at the point $w$ is propagated from the point $x$ by the retarded photon Green's function, while the step functions restrict the remaining integrations to the causal history encoded in the retarded response. The two terms in parentheses represent the order-$e^3$ corrections to the field generated by the induced current. The first term contains the polarization kernel inserted between the retarded photon propagators, whereas the second contains the three-point fermionic kernel with an internal photon line. Together they encode the electromagnetic dressing of the leading tadpole current in the external background. In contrast to the photon number density, this result is not reducible to a sum of squared transition amplitudes; it is an expectation value of the field operator and is naturally obtained within the nonequilibrium formulation.

The Schwinger-Dyson derivation in Appendix~\ref{sec:app_schwinger-dyson} provides an independent consistency check of Eq.~\eqref{eq:A3_full_final}.


\section{Conclusions} \label{sec:conclusions}

We have developed a perturbative description of radiative observables in QED with a prescribed external electromagnetic background capable of producing electron-positron pairs from the initial vacuum state. In this regime the vacuum persistence amplitude is nontrivial, the in- and out-vacua are inequivalent, and expectation values require a real-time nonequilibrium formulation. Using the Keldysh-Schwinger-Fradkin generating functional, we derived the mean photon number density and the mean electromagnetic field for a general spacetime-dependent background.

For the photon number density, we reproduced the leading order-$\alpha$ result, consisting of the vertex contribution associated with photon emission accompanying pair creation and the tadpole contribution generated by the induced vacuum current. We then evaluated the complete next-to-leading correction of order $\alpha^2$. The resulting expression contains all interference, loop, and induced-current terms arising at this order and is written in terms of exact in-solutions of the Dirac equation in the external field and the corresponding in-vacuum Green's functions. In the nonequilibrium formulation, vacuum-disconnected contributions are removed automatically.

We also obtained the same photon-number result by inserting the spectral decomposition of the identity operator in the in-Fock space. This direct representation rewrites the observable as a sum of squared transition amplitudes and makes the physical content of the different terms transparent. The vacuum-disconnected loops appearing in this approach cancel through optical-theorem relations generalized to an unstable vacuum. Agreement with the Keldysh-Schwinger-Fradkin calculation provides a nontrivial check of the final expression.

We further derived the mean electromagnetic field through order $e^3$. The final result has a manifestly causal structure in terms of retarded photon Green's functions and kernels describing the response and electromagnetic dressing of the induced current. Unlike the photon number density, the mean field is not reducible to a sum of squared transition amplitudes, which illustrates the necessity of the nonequilibrium approach for general expectation values in QED with unstable vacuum. We verified that the perturbative mean-field result satisfies the corresponding Schwinger-Dyson equations.

The formulas obtained here provide a general starting point for quantitative studies of higher-order vacuum radiation in strong external fields. The expressions derived in this work are unrenormalized and are intended as a systematic perturbative representation of the relevant expectation values before the implementation of a renormalization prescription. Since no assumption of spatial homogeneity, stationarity, or a specific field profile has been made, they can be applied to a broad class of pair-creating backgrounds once the exact fermionic modes or Green's functions are known. Natural extensions include the renormalization of the photon number density and mean electromagnetic field, numerical evaluations for realistic time-dependent backgrounds, comparisons of the different order-$\alpha^2$ channels, and self-consistent treatments of photon emission and the induced electromagnetic field.


\begin{acknowledgments}
We thank Prof.~S.~P.~Gavrilov and Prof.~V.~M.~Shabaev for valuable discussions. This work was supported by the Foundation for the Advancement of Theoretical Physics and Mathematics BASIS (Project No.~25-1-3-48-1).
\end{acknowledgments}


\appendix

\section{Useful identities and properties} \label{sec:app_properties}

The photon functions $D^{(\pm)}_{\mu \nu}$ have the spectral expansions
\begin{align}
D^{(-)}_{\mu \nu} (x, y) &= -i \sum_{\mathbf{k}, \lambda} f_{\mathbf{k} \lambda \mu} (x) f^*_{\mathbf{k} \lambda \nu} (y), \\
D^{(+)}_{\mu \nu} (x, y) &= i \sum_{\mathbf{k}, \lambda} f^*_{\mathbf{k} \lambda \mu} (x) f_{\mathbf{k} \lambda \nu} (y),
\end{align}
where $\sum_{\mathbf{k}, \lambda} \equiv \sum_{\lambda} (- g_{\lambda \lambda}) \int \! d\mathbf{k}$, with $\lambda = 0,1,2,3$. The physical photon modes are the transverse ones, $\lambda=1,2$. These functions satisfy
\begin{align}
D^{(+)}_{\nu \mu} (y, x) &= -D^{(-)}_{\mu \nu} (x, y), \\
\big [ D^{(+)}_{\mu \nu} (x, y) \big ]^* &= D^{(-)}_{\mu \nu} (x, y).
\end{align}
The photon spectral kernel
\begin{align}
D^{(0)}_{\mu \nu} (x, y) &= D^{(-)}_{\mu \nu} (x, y) + D^{(+)}_{\mu \nu} (x, y) \nonumber \\
{}&= -i  [ A_\mu (x), A_\nu (y)  ].
\end{align}
is real and obeys $D^{(0)}_{\nu \mu} (y, x) = -D^{(0)}_{\mu \nu} (x, y)$. The advanced and retarded photon Green's functions are
\begin{align}
D^{(\mathrm{A})}_{\mu \nu} (x, y) &= D_{\mu \nu} (x, y) - D^{(-)}_{\mu \nu} (x, y) \nonumber \\
{}&= -\theta(y^0 - x^0) D^{(0)}_{\mu \nu} (x, y), \\
D^{(\mathrm{R})}_{\mu \nu} (x, y) &= D_{\mu \nu} (x, y) + D^{(+)}_{\mu \nu} (x, y) \nonumber \\
{}&= \theta(x^0 - y^0) D^{(0)}_{\mu \nu} (x, y),
\end{align}
and they satisfy $D^{(\mathrm{A})}_{\nu \mu} (y, x) = D^{(\mathrm{R})}_{\mu \nu} (x, y)$. The matrix photon propagator in Eq.~\eqref{eq:DS_matrix_def_D} also has the properties
\begin{align}
D^{\gamma \beta}_{\nu \mu} (z_2, z_1) &= D^{\beta \gamma}_{\mu \nu} (z_1, z_2),\label{eq:prop_D_symm} \\
\big [ D^{\beta \gamma}_{\mu \nu} (z_1, z_2) \big ]^* &= -D^{\bar{\gamma} \bar{\beta}}_{\nu \mu} (z_2, z_1) = -D^{\bar{\beta} \bar{\gamma}}_{\mu \nu} (z_1, z_2). \label{eq:Dconj_gen}
\end{align}
Here $\bar{\beta} = 2$ for $\beta = 1$ and $\bar{\beta} = 1$ for $\beta = 2$. Finally, we note that $D_{\mu \nu} + D^{(+)}_{\mu \nu} - D^{(-)}_{\mu \nu} + \overline{D}_{\mu \nu} = 0$.

For the fermion functions, the spectral expansions are
\begin{align}
S^{(-)} (x, y) &= i \sum_{n} {}_+ \varphi_n (x) {}_+ \overline{\varphi}_n (y), \label{eq:Sminus} \\
S^{(+)} (x, y) &= i \sum_{n} {}_- \varphi_n (x) {}_- \overline{\varphi}_n (y). \label{eq:Splus}
\end{align}
The matrix fermion propagator obeys
\begin{equation}
\gamma^0 \big [ S^{\lambda_1 \lambda_2} (x, y) \big ]^\dagger \gamma^0 = - S^{\bar{\lambda}_2 \bar{\lambda}_1} (y,x). \label{eq:Sconj_gen}
\end{equation}
%


\section{Schwinger-Dyson equations and consistency check} \label{sec:app_schwinger-dyson}

In this appendix we derive the Schwinger-Dyson equations that follow from the generating functional introduced in Sec.~\ref{sec:setup}. We then rewrite them as integral equations for the exact photon and fermion Green's functions and use their perturbative expansion as a consistency check of the third-order mean electromagnetic field.

\subsection{Schwinger-Dyson equations}

Let
\begin{equation}
\Box_{\beta \gamma} \equiv \begin{pmatrix}
\Box & 0 \\
0 & -\Box
\end{pmatrix}_{\beta \gamma}.
\end{equation}
All differential operators in this appendix act on the $x$ variable of the corresponding function. For the causal photon Green's function, $\Box D_{\mu \nu} (x,y) = g_{\mu \nu} \delta (x-y)$, and the matrix relation is
\begin{equation}
\Box_{\beta \gamma} D_{\mu \nu}^{\gamma \delta} (x, y) = g_{\mu \nu} \delta (x-y) \delta_{\beta \delta}.
\end{equation}
The corresponding Dirac operator is
\begin{equation}
\mathcal{D}^{\alpha \lambda} (\mathcal{A}) \equiv \begin{pmatrix}
\mathcal{D} (\mathcal{A}) & 0 \\
0 & -\mathcal{D} (\mathcal{A})
\end{pmatrix}_{\alpha \lambda},
\end{equation}
where
\begin{equation}
\mathcal{D} (\mathcal{A}) = [i \partial_\mu - e \mathcal{A}_\mu (x)] \gamma^\mu - m.
\end{equation}
The fermion matrix propagators obey
\begin{equation}
\mathcal{D}^{\beta \alpha} (\mathcal{A}) S^{\alpha \lambda} (x, y) = - \delta_{\beta \lambda} \delta (x-y).
\end{equation}

For connected Green's functions set
\begin{equation}
W = i \ln Z.
\end{equation}
The required one- and two-point functions are defined as
\begin{align}
\alpha_\mu^\beta (x) &\equiv \frac{\delta W}{\delta I^\mu_\beta (x)} \bigg |_{\eta=\overline{\eta}=0}, \label{eq:alpha_def} \\
\mathbb{S}^{\alpha \lambda}_{l l'} (x, y) &\equiv \frac{\delta^2 W}{\delta \overline{\eta}_{\alpha l} (x) \delta \eta_{\lambda l'} (y)} \bigg |_{\eta=\overline{\eta}=0}, \label{eq:S_full_def} \\
\mathbb{D}^{\beta \gamma}_{\mu \nu} (x, y) &\equiv \frac{\delta^2 W}{\delta I^\mu_\beta (x) \delta I^\nu_\gamma (y)} \bigg |_{\eta=\overline{\eta}=0}.
\end{align}
Keeping the photon sources nonzero for the moment, one directly obtains
\begin{equation}
\Box_{\lambda \beta} \alpha_\mu^\beta (x) = g_{\mu \nu} I^\nu_\lambda (x) + ie g_{\mu \nu} \operatorname{Tr} \big [ \gamma^\nu \mathbb{S}^{\lambda \lambda} (x, x) \big ],
\label{eq:box_alpha}
\end{equation}
where no summation over $\lambda$ is implied. Next, by straightforward algebra, one obtains
\begin{align}
\Box_{\lambda \beta} \mathbb{D}^{\beta \alpha}_{\mu \nu} (x, y) &= g_{\mu \nu} \delta_{\lambda \alpha} \delta (x-y) \nonumber \\
{}&+ ie \operatorname{Tr} \bigg [ \gamma_\mu \frac{\delta \mathbb{S}^{\lambda \lambda} (x, x)}{\delta I^\nu_\alpha (y)} \bigg ].
\label{eq:boxD}
\end{align}
For the fermion sector, we define
\begin{multline}
\mathcal{D}^{\alpha \lambda} (\mathcal{A} + i\delta_I + \alpha) \\
{}\equiv \begin{pmatrix}
\mathcal{D} \Big ( \mathcal{A} + i \frac{\delta}{\delta I_1} + \alpha^1 \Big ) & 0 \\
0 & -\mathcal{D} \Big ( \mathcal{A} - i \frac{\delta}{\delta I_2} - \alpha^2 \Big )
\end{pmatrix}_{\alpha \lambda} \label{eq:D_AIalpha}
\end{multline}
and find
\begin{equation}
\mathcal{D}^{\beta \alpha} (\mathcal{A} + i\delta_I + \alpha) \mathbb{S}^{\alpha \lambda} (x, y) = -\delta_{\beta \lambda} \delta (x - y).
\label{eq:SD_3}
\end{equation}
Equations~\eqref{eq:box_alpha}, \eqref{eq:boxD}, and \eqref{eq:SD_3} form the Schwinger-Dyson system in the nonequilibrium formulation. We now cast this system into an integral form better suited for perturbation theory.

\subsection{System of integral equations and the mean field}

To eliminate the functional derivatives $\delta/\delta I$, we introduce the polarization and mass operators
\begin{align}
\Pi^{\mu \nu}_{\lambda \gamma} (x, y) &\equiv -ie^2 \operatorname{Tr} \bigg [ \gamma^\mu \int \! d^4z \int \! d^4z' \, \mathbb{S}^{\lambda \alpha} (x, z) \nonumber \\
{}&\times \Gamma^\nu_{\alpha \beta \gamma} (z, z', y) \mathbb{S}^{\beta \lambda} (z', x) \bigg ], \\
\Sigma^{\lambda \gamma} (x,y) &\equiv -ie^2 \gamma^\mu \int \! d^4z \int \! d^4z' \, \mathbb{S}^{\lambda \alpha} (x, z) \nonumber \\
{}&\times \Gamma^\nu_{\alpha \gamma \beta} (z, y, z') \mathbb{D}^{\beta \lambda}_{\nu \mu} (z', x).
\end{align}
Here $\Gamma$ denotes the full vertex operator. These definitions imply
\begin{align}
ie \operatorname{Tr} \bigg [ \gamma^\mu \frac{\delta \mathbb{S}^{\lambda \lambda} (x, x)}{\delta I^\nu_\gamma (y)} \bigg ] &= \int \! d^4z \, \Pi^{\mu \rho}_{\lambda \alpha} (x, z) \mathbb{D}^{\alpha \gamma}_{\rho \nu} (z, y), \\
ie \gamma^\mu \, \frac{\delta \mathbb{S}^{\lambda \gamma} (x, y)}{\delta I^\mu_\lambda (x)} &= \int \! d^4z \, \Sigma^{\lambda \beta} (x, z) \mathbb{S}^{\beta \gamma} (z, y).
\end{align}
The system of equations takes the form
\begin{align}
\Box_{\lambda \beta} \alpha^{\mu \beta} (x) &= I^\mu_\lambda (x) + ie \operatorname{Tr} \big [ \gamma^\mu \mathbb{S}^{\lambda \lambda} (x, x) \big ], \label{eq:alphaDS_alpha}\\
\Box_{\lambda \beta} \mathbb{D}^{\beta \alpha}_{\mu \nu} (x, y) &= g_{\mu \nu} \delta_{\lambda \alpha} \delta (x-y) \nonumber \\
{}&+ g_{\mu \sigma} \int \! d^4z \, \Pi^{\sigma \rho}_{\lambda \gamma} (x, z) \mathbb{D}^{\gamma \alpha}_{\rho \nu} (z, y),\label{eq:alphaDS_D} \\
\mathcal{D}^{\beta \lambda} (\mathcal{A} + \alpha) \mathbb{S}^{\lambda \gamma} (x, y) &= -\delta_{\beta \gamma} \delta (x - y) \nonumber \\
{}&+ \int \! d^4z \, \Sigma^{\beta \alpha} (x, z) \mathbb{S}^{\alpha \gamma} (z, y). \label{eq:alphaDS_S}
\end{align}
This system can be solved perturbatively and therefore provides an alternative construction of the full Green's functions. We now apply it to the mean electromagnetic field.

For the vacuum initial state and vanishing photon source, $I=0$, the mean field~\eqref{eq:Amean_Z_full} is
\begin{equation}
\langle A_\mu (x) \rangle = \alpha_\mu^1 (x) = -\alpha_\mu^2 (x).
\label{eq:A_alpha}
\end{equation}
From Eq.~\eqref{eq:alphaDS_alpha} it follows that
\begin{equation}
\Box \langle A_\mu (x) \rangle = ie \operatorname{Tr} \big [ \gamma_\mu \mathbb{S}^{11} (x, x) \big ].
\label{eq:Amean_eq}
\end{equation}
At zeroth order in the fine-structure constant, $\mathbb{S}^{11} (x, y)$ reduces to the causal propagator $S(x,y)$, and Eq.~\eqref{eq:Amean_eq} gives the leading mean field~\eqref{eq:Amean_1_final}. With retarded boundary conditions, the all-order solution is
\begin{equation}
\langle A_\mu (x) \rangle = ie \int \! d^4 x' \, D^{(\mathrm{R})}_{\mu \nu} (x, x') \operatorname{Tr} \big [ \gamma^\nu \mathbb{S}^{11} (x', x') \big ].
\label{eq:A_S_full}
\end{equation}

The order-$e^3$ mean field requires the order-$e^2$ correction to the propagator $\mathbb{S}^{\lambda \alpha} (x, y)$. Denoting this correction by $\mathbb{S}^{\lambda \alpha}_2 (x, y)$, we find
\begin{equation}
\langle A_\mu (x) \rangle^{(3)} = ie \int \! d^4 x' \, D^{(\mathrm{R})}_{\mu \nu} (x, x') \operatorname{Tr} \big [ \gamma^\nu \mathbb{S}^{11}_2 (x', x') \big ].
\label{eq:Amean_3}
\end{equation}
We next derive the equation obeyed by $\mathbb{S}^{11}_2 (x, y)$. To second order in $e$, Eq.~\eqref{eq:alphaDS_S} gives
\begin{multline}
\mathcal{D}^{\beta \lambda} (\mathcal{A}) \mathbb{S}^{\lambda \gamma}_2 (x, y) - e \gamma^\mu \alpha_{1\mu}^\beta (x) S^{\beta \gamma} (x, y) \\
{}= \int \! d^4z \, \Sigma_2^{\beta \alpha} (x, z) S^{\alpha \gamma} (z, y),
\end{multline}
where $\alpha_{1\mu}^\beta (x)$ is the leading-order part of $\alpha_{\mu}^\beta (x)$, determined by Eq.~\eqref{eq:A_alpha}. Setting $\beta = \gamma = 1$ yields
\begin{multline}
\mathcal{D} (\mathcal{A}) \mathbb{S}^{11}_2 (x, y) - e \gamma^\mu \langle A_\mu (x) \rangle^{(1)} S(x,y) \\
{}= \int \! d^4z \, \Sigma_2^{1 \alpha} (x, z) S^{\alpha 1} (z, y).
\end{multline}
The leading-order mass operator is given by
\begin{equation}
\Sigma_2^{\lambda \gamma} (x,y) = -ie^2 \gamma^\mu S^{\lambda \gamma} (x, y) \gamma^\nu D^{\gamma \lambda}_{\nu \mu} (y, x).
\end{equation}
It follows that
\begin{widetext}
\begin{equation}
\mathcal{D} (\mathcal{A}) \mathbb{S}^{11}_2 (x, y) = e \gamma^\mu \langle A_\mu (x) \rangle^{(1)} S(x,y) - ie^2 \int \! d^4z \, \gamma^\mu S^{1 \alpha} (x, z) \gamma^\nu D^{\alpha 1}_{\nu \mu} (z, x) S^{\alpha 1} (z, y).
\label{eq:S2_eq}
\end{equation}
This differential equation alone does not uniquely determine the propagator correction $\mathbb{S}^{11}_2 (x, y)$. We therefore compute the correction directly from the definition~\eqref{eq:S_full_def}:
\begin{equation}
\mathbb{S}^{\lambda \alpha}_{2 ll'} (x, y) = i \, \frac{\delta^2 Z^{(2)}}{\delta \overline{\eta}_{\lambda l} (x) \delta \eta_{\alpha l'} (y)} \bigg |_{I=\eta=\overline{\eta}=0}.
\label{eq:S2_direct}
\end{equation}
We then verify that the explicit result satisfies Eq.~\eqref{eq:S2_eq} and, after substitution into Eq.~\eqref{eq:Amean_3}, reproduces Eq.~\eqref{eq:A3_full_final}.

Evaluating the functional derivatives in Eq.~\eqref{eq:S2_direct} yields
\begin{align}
\mathbb{S}^{\lambda \alpha}_{2} (x, y) &= -e \int \! d^4z_1 \int \! d^4z_2 \, j^\mu_\mathrm{in} (z_1) D^{\lambda_1 \lambda_2}_{\mu \nu} (z_1, z_2) S^{\lambda \lambda_2} (x, z_2) \gamma^\nu S^{\lambda_2 \alpha} (z_2, y) \nonumber \\
{}&+ ie^2 \int \! d^4z_1 \int \! d^4z_2 \, D^{\lambda_1 \lambda_2}_{\mu \nu} (z_1, z_2) S^{\lambda \lambda_1} (x, z_1) \gamma^\mu S^{\lambda_1 \lambda_2} (z_1, z_2) \gamma^\nu S^{\lambda_2 \alpha} (z_2, y).
\end{align}
Diagrammatically, these two terms are the second-order radiative corrections generated by the leading induced current and by an internal photon line. After summing over $\lambda_1$ and $\lambda_2$, the component with $\lambda = \alpha = 1$ becomes
\begin{align}
\mathbb{S}^{11}_{2} (x, y) &= -e \int \! d^4z_1 \int \! d^4z_2 \, j^\mu_\mathrm{in} (z_1) D^{(\mathrm{R})}_{\nu \mu} (z_2, z_1) \nonumber \\
{}&\times \big [ S (x, z_2) \gamma^\nu S (z_2, y) + S^{(+)} (x, z_2) \gamma^\nu S^{(-)} (z_2, y) \big ] \nonumber \\
{}&+ ie^2 \int \! d^4z_1 \int \! d^4z_2 \, \big [ D_{\mu \nu} (z_1, z_2) S (x, z_1) \gamma^\mu S(z_1, z_2) \gamma^\nu S (z_2, y) \nonumber \\
{}&- D^{(+)}_{\mu \nu} (z_1, z_2) S (x, z_1) \gamma^\mu S^{(+)} (z_1, z_2) \gamma^\nu S^{(-)} (z_2, y) \nonumber \\
{}&+ D^{(-)}_{\mu \nu} (z_1, z_2) S^{(+)} (x, z_1) \gamma^\mu S^{(-)}(z_1, z_2) \gamma^\nu S (z_2, y) \nonumber \\
{}&- \overline{D}_{\mu \nu} (z_1, z_2) S^{(+)} (x, z_1) \gamma^\mu \overline{S}(z_1, z_2) \gamma^\nu S^{(-)} (z_2, y) \big ].
\label{eq:S2_explicit}
\end{align}
Acting with $\mathcal{D} (\mathcal{A})$ on Eq.~\eqref{eq:S2_explicit}, we find
\begin{align}
\mathcal{D} (\mathcal{A}) \mathbb{S}^{11}_{2} (x, y) &= e \int \! d^4z \, j^\mu_\mathrm{in} (z) D^{(\mathrm{R})}_{\nu \mu} (x, z) \gamma^\nu S (x, y) \nonumber \\
{}&- ie^2 \int \! d^4z \, \big [ D_{\mu \nu} (x, z) \gamma^\mu S(x, z) \gamma^\nu S (z, y) - D^{(+)}_{\mu \nu} (x, z) \gamma^\mu S^{(+)} (x, z) \gamma^\nu S^{(-)} (z, y) \big ].
\end{align}
This is precisely the right-hand side of Eq.~\eqref{eq:S2_eq}.

Finally, substituting Eq.~\eqref{eq:S2_explicit} into Eq.~\eqref{eq:Amean_3} and following the same algebraic steps as in the main-text derivation gives Eq.~\eqref{eq:A3_full_final}. The perturbative solution of the Schwinger-Dyson equations therefore reproduces the third-order mean field obtained from the generating functional.

\end{widetext}


\end{document}